\documentclass{evolang}
\usepackage{graphicx}
\usepackage{url}
\usepackage{authblk}

\usepackage{float}
\usepackage{xcolor, soul}
\sethlcolor{green}
\usepackage{appendix}
\usepackage{booktabs}
\usepackage{longtable}
\usepackage{rotating}
\usepackage{multirow}

\usepackage[labelsep=space]{caption}
\usepackage{subcaption}

\newcommand{\eg}{\emph{e.g.}}

\newcommand{\fref}[1]{Figure~\ref{#1}}

\begin{document}

\title{The emergence of numerical representations\\ in communicating artificial agents}
\author[*1]{Daniela Mihai}
\author[2]{Lucas Weber}
\author[2]{Francesca Franzon}
\affil[*]{Corresponding Author: a-d.mihai@soton.ac.uk}
\affil[1]{School of Electronics and Computer Science, University of Southampton, United Kingdom}
\affil[2]{Department of Translation and Language Sciences, Universitat Pompeu Fabra, Barcelona, Spain}

\maketitle

\abstracts{
Human languages provide efficient systems for expressing numerosities, but whether the sheer pressure to communicate is enough for numerical representations to arise in artificial agents, and whether the emergent codes resemble human numerals at all, remains an open question. We study two neural network-based agents that must communicate numerosities in a referential game using either discrete tokens or continuous sketches, thus exploring both symbolic and iconic representations. Without any pre-defined numeric concepts, the agents achieve high in-distribution communication accuracy in both communication channels and converge on high-precision symbol–meaning mappings. However, the emergent code is non-compositional: the agents fail to derive systematic messages for unseen numerosities, typically reusing the symbol of the highest trained numerosity (discrete), or collapsing extrapolated values onto a single sketch (continuous). We conclude that the communication pressure alone suffices for precise transmission of learned numerosities, but additional pressures are needed to yield compositional codes and generalisation abilities.}

\section{Communicating and representing numerosities}
Number words, such as \emph{two, ten, sixteen}, allow humans to encode any numerosity up to infinity, relying on finite memory. 
Their encoding efficiency lies in several properties. First, they exhibit \textbf{precision}: each word maps one-to-one onto a numerosity; this mapping can be extended indefinitely through extrapolation, assigning a new word to each added element.
They also follow transparent \textbf{compositional rules}, reducing the number of forms to memorise: for example, knowing the words for units and tens allows speakers to interpolate intermediate numerosities. 
These properties enable learnability, making concept–word mappings easy to infer and \textbf{generalise} to novel words and novel numerosities: in humans, understanding the structure of number words unlocks precise representation of any numerical concept~\cite{barner2017language}, whereas without number words, quantity representation has been argued to be only approximate.
Finally, the form of number words is \textbf{arbitrary}, as no property of the word suggests the numerosity they refer to; however, as commonly in language, most frequently occurring number words are shorter~\cite{pajot2025compositional}. 
However, written representations of numbers vary in arbitrariness and can start as iconic representations, where the visual properties of number symbols directly reproduce the numerosity to encode~\cite{overmann2025cultural}.

Although numerical systems emerged independently in different cultures~\cite{coolidge2012numerosity}, number words across languages largely share these properties~\cite{wals-131}, suggesting that they reflect optimisation to similar encoding pressures~\cite{xu2020numeral,denic2024recursive}. 

Language evolution theories propose that structured systems arise under pressures for transmission, economy, and generalisation~\cite{kirby2008cumulative}. 
Across cultures, numbers originated as a cultural invention for unambiguous communication, used to record stored goods and track quantities in transactions~\cite{menninger2013number}. 
We therefore ask whether communicative pressures shape number words and, in turn, enable the representation of number concepts. 

Specifically, we ask (i) whether the properties observed in natural language numerals also arise in a code self-optimised through communication, (ii) which aspects of communication support them and (iii) whether the resulting code supports generalisation to unseen numerosities, bootstrapping their conceptual representation. To this end, we implement a referential game ~\cite{lewis1969convention,ReviewEmeComm,brandizzi2023toward}, in which a communication protocol is shaped solely through collaborative interaction between two neural-network agents without prior knowledge of numerals. We test whether the pressure to communicate is sufficient for agents to develop a code, and whether it shows precision, compositionality, generalisation, and arbitrariness as emerging properties. We also assess whether frequency and learning strategy (interpolation vs extrapolation) affect communication success and code structure.

\section{Environmental Setup}

\begin{figure*}[ht]
    \begin{subfigure}{0.65\linewidth}
    \includegraphics[width=1\textwidth]{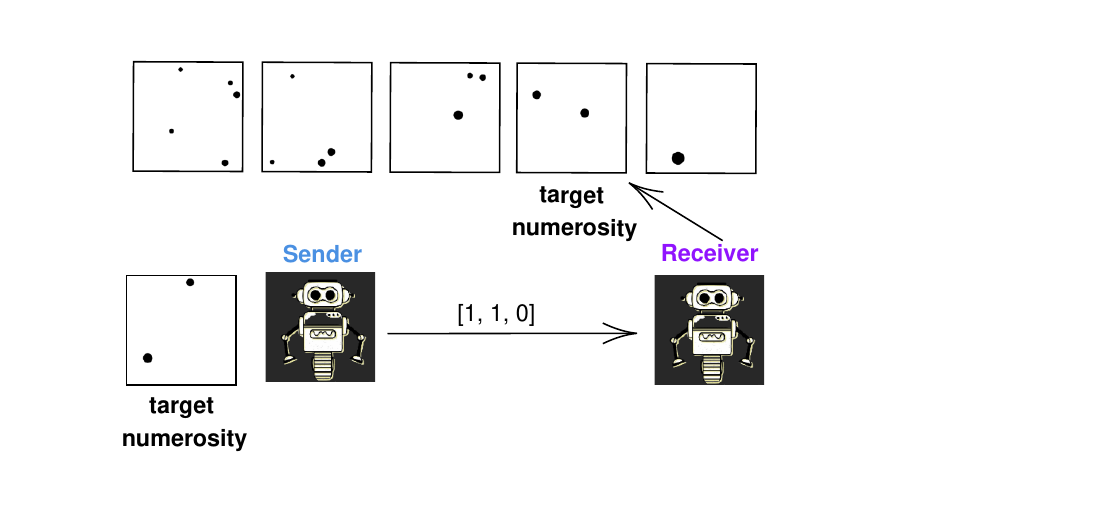}
    \vspace{-3em}
    \hspace{-3cm}\caption{}
    \label{fig:referential-game}
    \end{subfigure}
    \hspace{-1.4cm}
    \begin{subfigure}{0.33\linewidth}
    \includegraphics[width=1.15\textwidth]{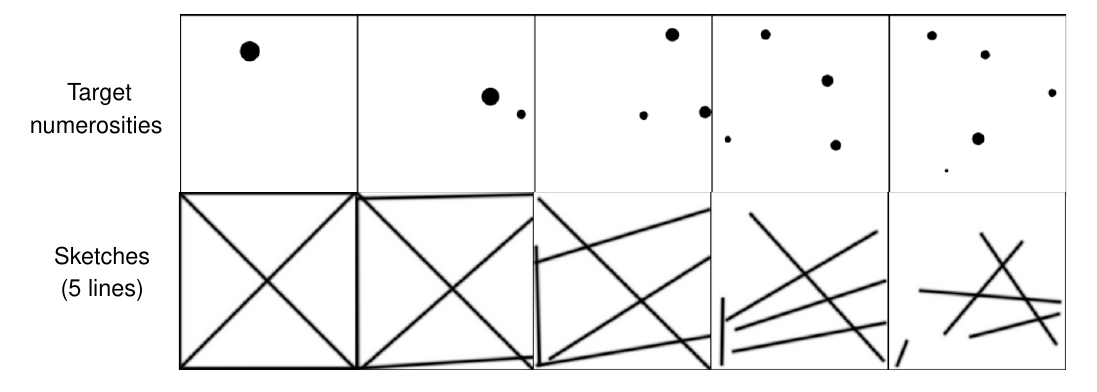}
    \vspace{0.5em}
    \caption{}
    \label{subfig:num_sketches_examples}
\end{subfigure}
\caption{\footnotesize (a) A referential game: the Receiver has to guess, based on a Sender's discrete message, the correct target out of 5 possible numerosities. Notice that although the target is also 2, it is not the same instance as seen by the Sender. (b) Examples of target numerosities $\{1,2,3,4,5\}$ and their 5-line corresponding sketches.}
\end{figure*}

\paragraph{Communication Task.}
We used a referential game setup (Fig.~\ref{fig:referential-game}), where two neural-network agents communicate about numerosities. In each trial, a sender agent is presented with an image displaying a certain number of dots and sends a message to a receiver agent. The receiver, based on the message, must identify the \emph{target} image showing the same number of dots, among distractors with different numerosities. Upon target selection, both agents receive feedback; in this way, the communication protocol emerges while solving the task. The setup has a \emph{training} phase, where agents receive feedback and therefore develop their code, and a \emph{generalisation} phase, where they communicate about novel numerosities.

\vspace{-0.7em}

\paragraph{Stimuli.} 
We designed the stimuli to ensure that agents rely only on numerosity, preventing confounds of other visual properties, like shape, colour, area or perimeter occupied by the dots.
Stimuli images are 256 x 256 pixels in size, each depicting a different number of black dots on a white background (e.g. Fig.~\ref{subfig:num_sketches_examples}). 
In order to make information on numerosity independent of other visual variables such as the area occupied by black pixels, we constrain the total black area occupied by dots to 5-10\% of the canvas for all numerosities. 

\vspace{-0.8em}
\paragraph{Conditions.} For the purpose of this work, the main training set is mostly limited to numerosities between 1 and 5 (although we experimented with larger training sets, such as 1 to 20, and observed similar results). The extrapolation/interpolation evaluation can include one other numerosity. The exact training and testing numerosities are specified for each experiment. To test if representational abilities of numerosity emerge independently of the low-level visual properties, we test two target setups, \emph{Same} and \emph{Diff}, which indicate whether the agents have the exact \emph{same} target image, or two \emph{different} instances from the same numerosity class.

\vspace{-0.5em}
\paragraph{Agents.} Both agents use a shared pretrained ViT encoder for processing the dot images \cite{radford2021learning}. We model two types of communication channels. In the \textbf{discrete} setup, the sender is equipped with an LSTM encoder \cite{hochreiter1997long}, which outputs a message as a sequence of discrete tokens chosen over a predefined vocabulary; the receiver's LSTM reads the message and accordingly selects an image as a possible target. In the \textbf{continuous} setup, the sender draws a fixed number of straight lines on a blank canvas, i.e. a sketch \cite{mihai2021learningtodraw}. The receiver reads the sketch through the same ViT encoder used for the candidate target images. A multi-class hinge loss is employed to maximise the probability that the correct target image is selected. Training happens on a set of numerosities as specified per experiment, and the testing is both in-range and on unseen values. 

\vspace{-0.5em}
\paragraph{Metrics.} We assess communication success via task accuracy, defined as the proportion of trials in which the receiver selects the numerosity presented to the sender. Additionally, we assess the properties of the emerging communication code: \textit{Efficiency} via message length (tokens) or sketch strokes/span; \textit{Precision} as conditional entropy of numerosity given message (sketches are discretised via clustering); \textit{Generalisation} by accuracy on unseen numerosities and where their messages map among seen classes; and \textit{Arbitrariness} by skewing training frequencies and checking if more frequent numbers get shorter or simpler codes.

\section{Results}
\begin{figure}[ht]
\centering
\begin{subfigure}{0.32\linewidth}
    \includegraphics[width=\linewidth]{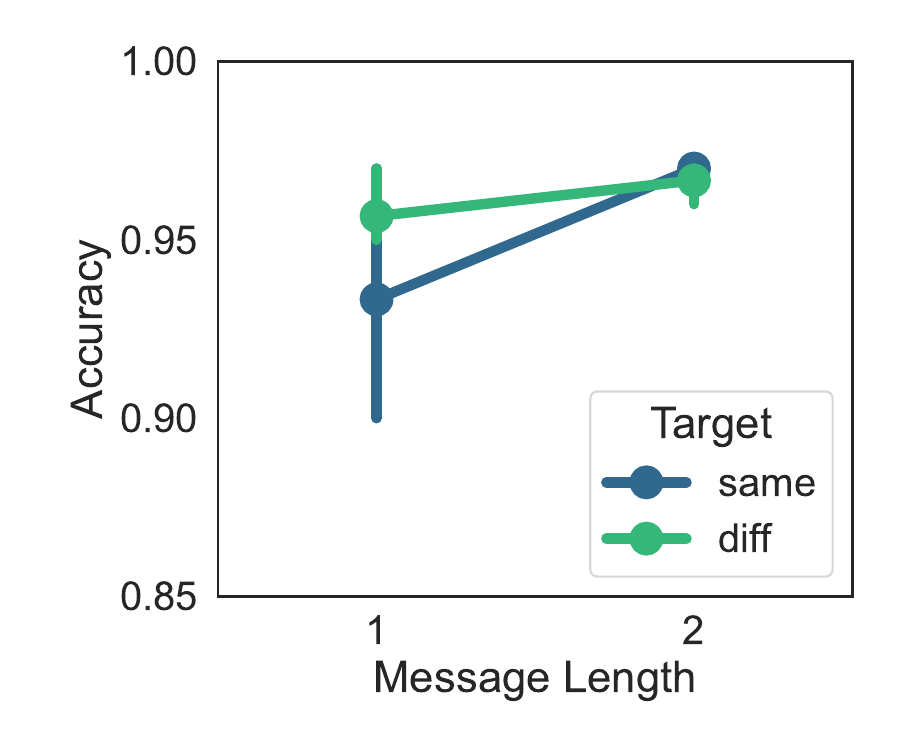}
    \caption{}
    \label{subfig:pointplot_accuracy_same_v_diff}
\end{subfigure}
\hfill
\begin{subfigure}{0.32\linewidth}
    \includegraphics[width=\linewidth]{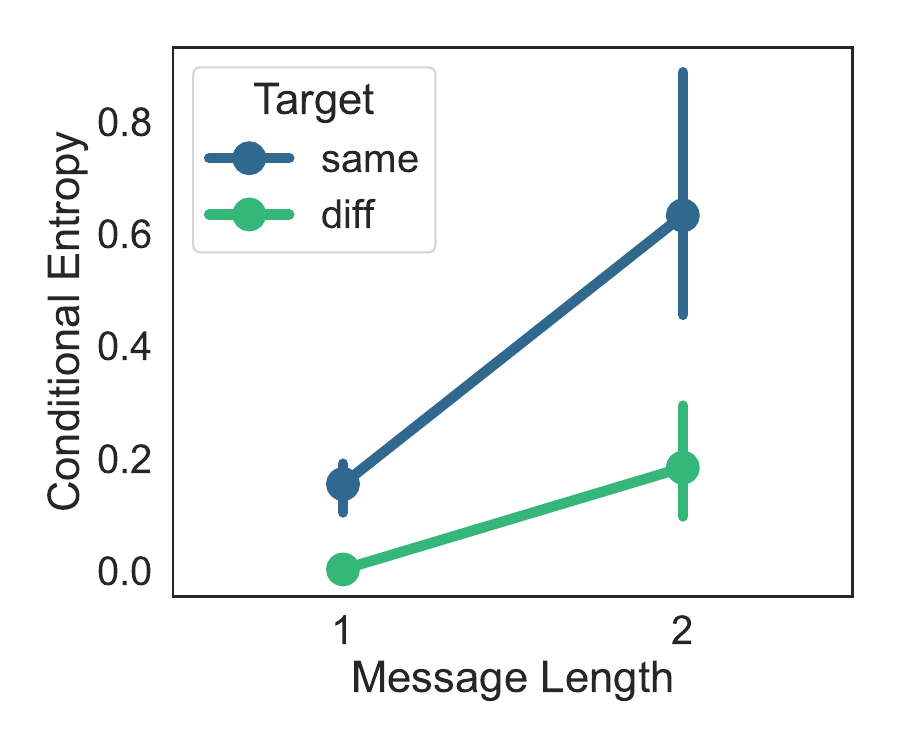}
    \caption{\footnotesize}
    \label{subfig:pointplot_conditional_e_same_v_diff}
\end{subfigure}
\hfill
\begin{subfigure}{0.32\linewidth}
  {\includegraphics[width=\linewidth]{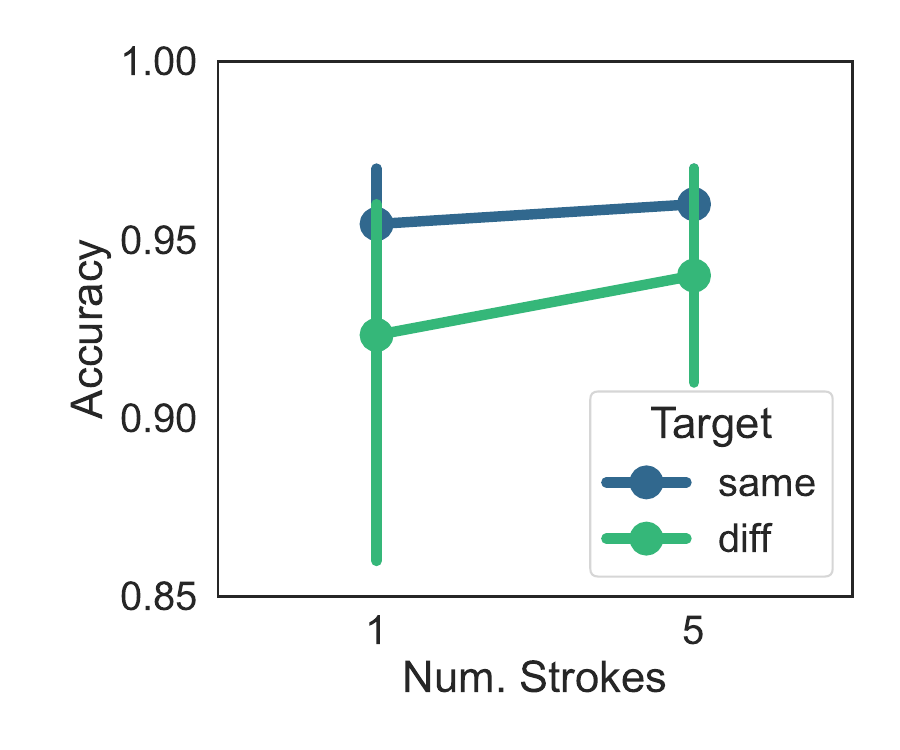}}
    \caption{\footnotesize}
    \label{subfig:pointplot_sketch_acc_same_vs_diff}
\end{subfigure}
\caption{\footnotesize (a) Discrete communication accuracy and (b) message entropy, and (c) continuous (i.e. sketch-based) communication accuracy for setups involving identical target pictures (same - blue line) and different pictures showing the same numerosity (diff - green line). Model trained and tested on numerosities 1 to 5. Results averaged over 3 seeds.}
\label{fig:same_v_diff}
\end{figure}

\begin{figure*}[hb!]
\centering
\begin{subfigure}{0.32\linewidth}
    \includegraphics[width=1.\linewidth]{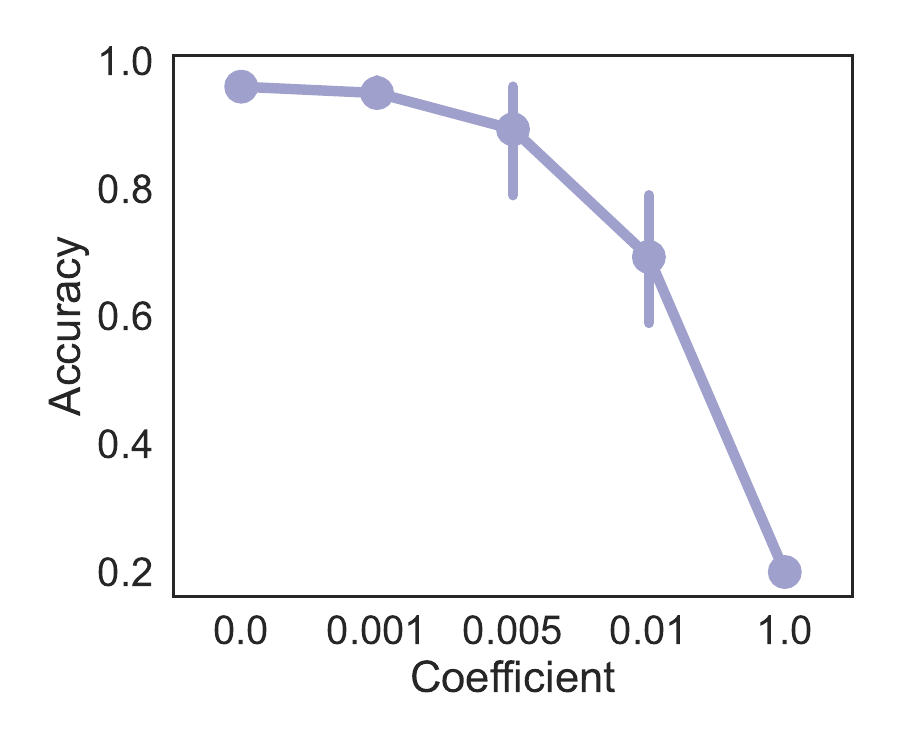}
\caption{}
\label{subfig:accuracy_msgregularization_discrete}
\end{subfigure}
\hfill
    \begin{subfigure}{0.32\linewidth}
    \includegraphics[width=1.\linewidth]{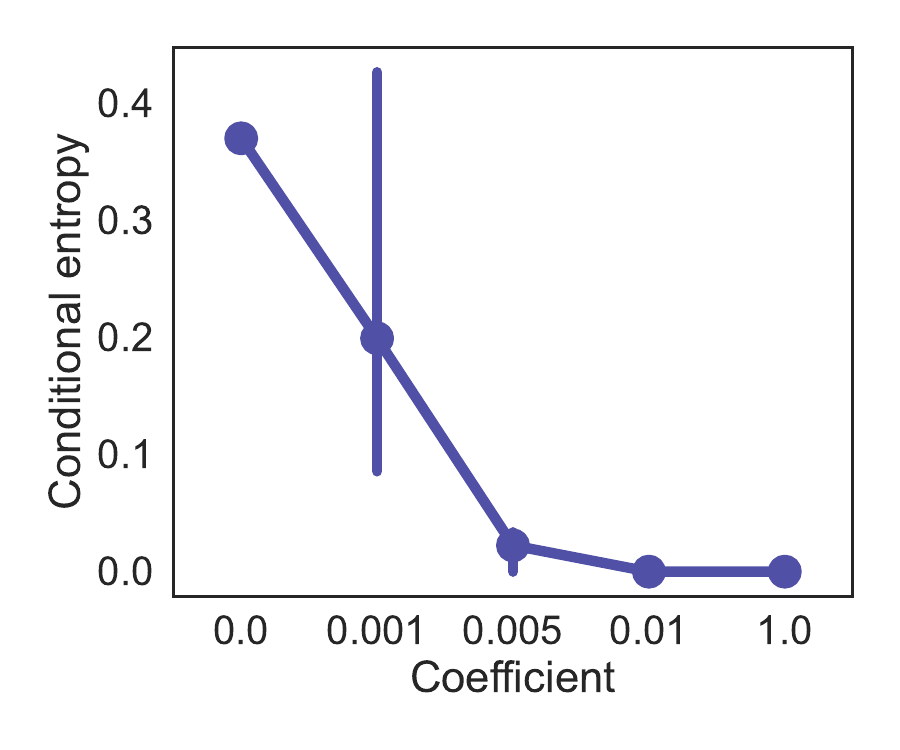}
\caption{}
\label{subfig:cond_entropy_msgregularization_discrete}
\end{subfigure}
\hfill
\begin{subfigure}{0.32\linewidth}
    \includegraphics[width=1.\linewidth]{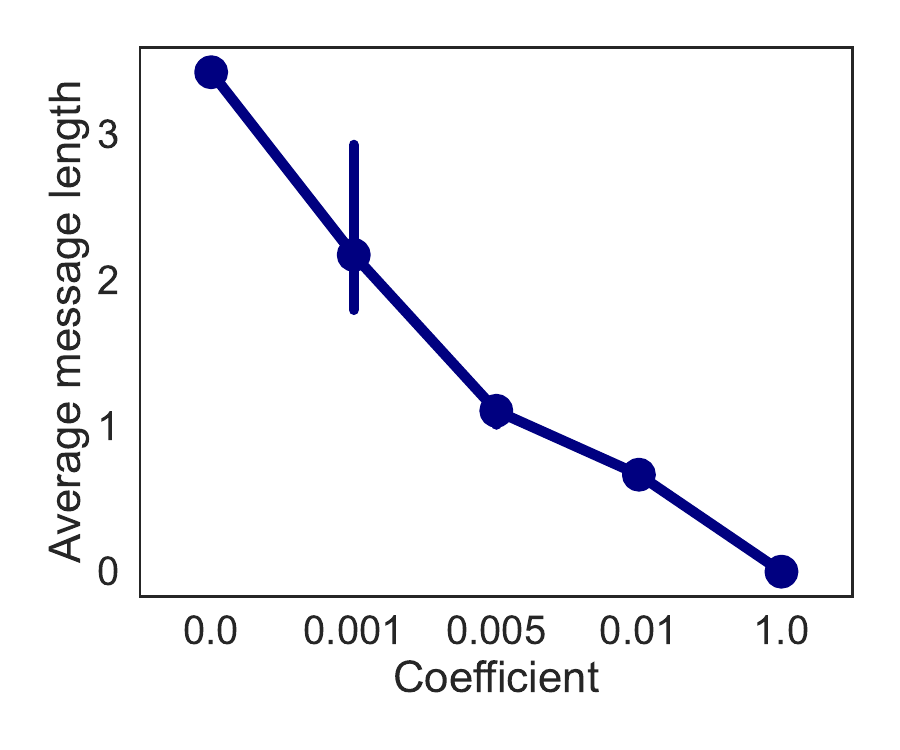}
\caption{}
\label{subfig:msg_length_msgregularization_discrete}
\end{subfigure}
\vspace{-0.5em}
\caption{\footnotesize Discrete Communication - Agent messaging behaviour when longer messages are penalised by regularising the game loss. Model trained on numerosities $\{1,2,3,4,5\}$, with a variable message length allowed up to \textbf{5} tokens, and $vocabulary=3$. (a) Accuracy (b) conditional entropy of the mapping, and (c) average message length for different regularisation coefficients.}
\label{fig:discrete-regularization}
\end{figure*}

\begin{figure}[htb]
\centering
\begin{subfigure}{0.3\linewidth}
    \includegraphics[width=\linewidth]{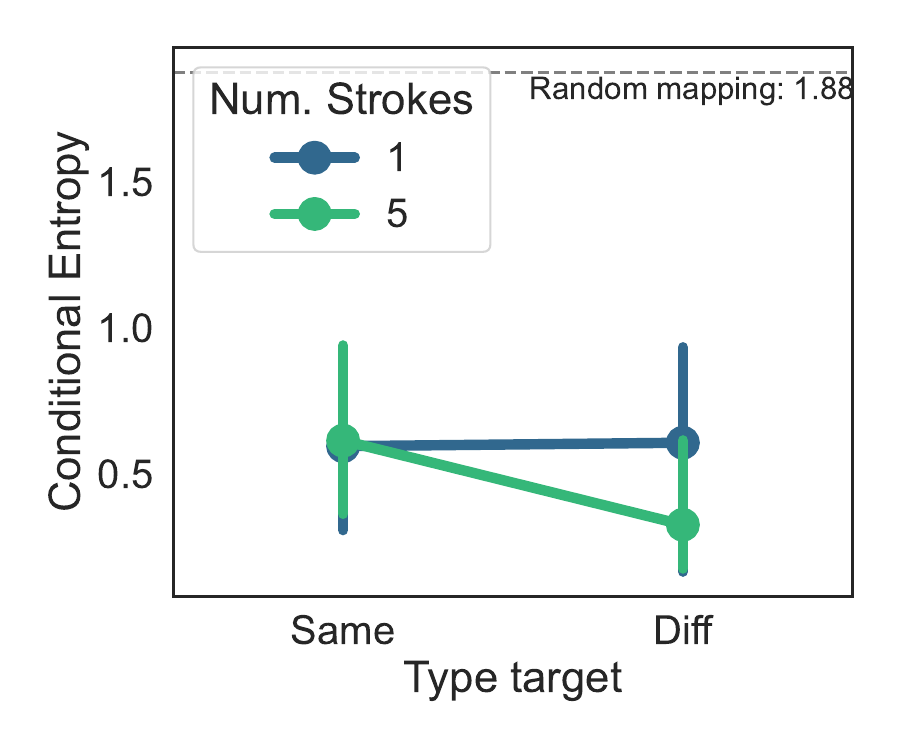}
    \caption{}
    \label{subfig:point_plot_conditional_entropy_same_v_diff_1v5_strokes}
\end{subfigure}
\begin{subfigure}{0.3\linewidth}
    \includegraphics[width=\linewidth]{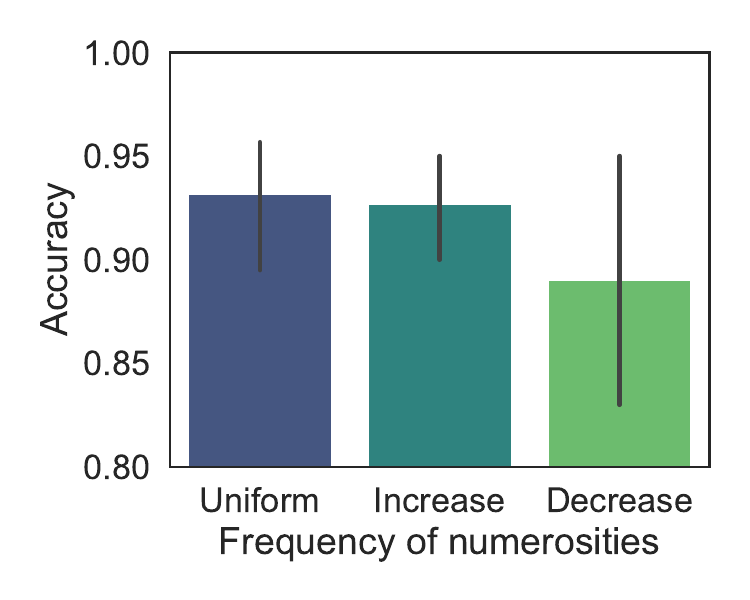}
    \caption{}
\label{subfig:point_plot_accucary_increase_v_decrease}
\end{subfigure}
\begin{subfigure}{0.3\linewidth}
    \includegraphics[width=\linewidth]{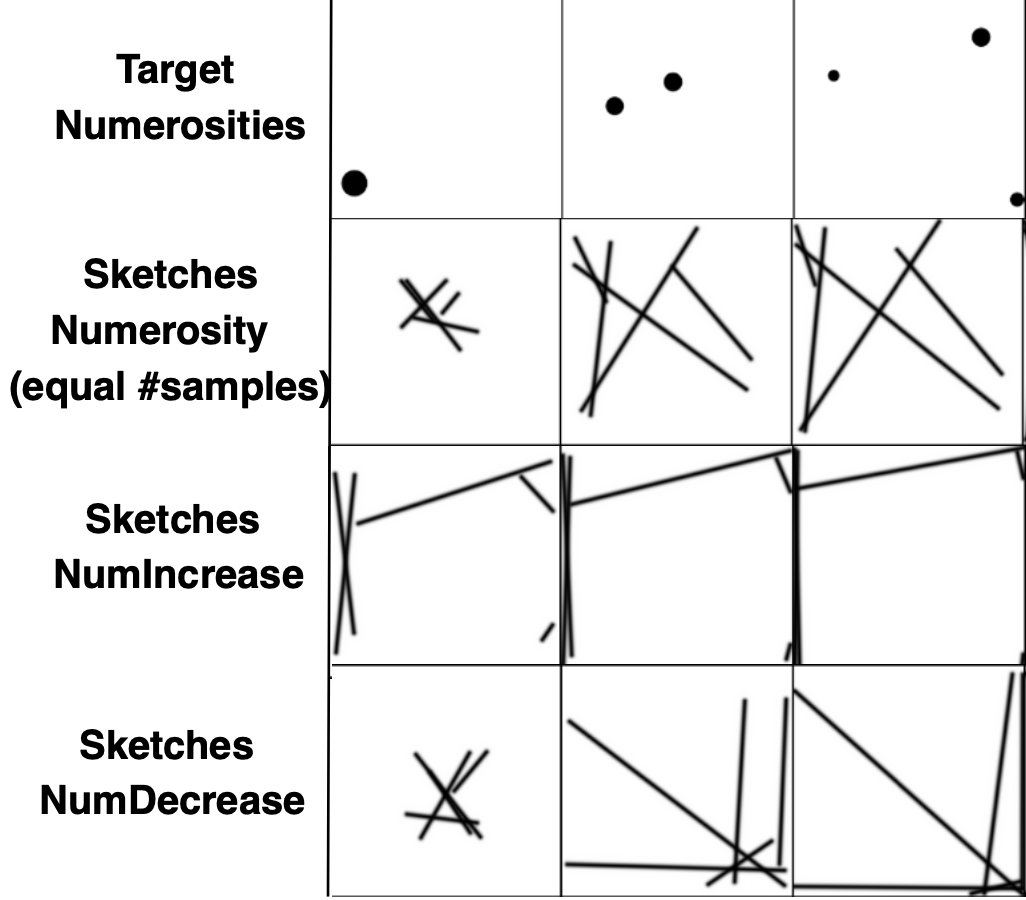}
    \caption{}
\label{subfig:sketch_examples_inc_dec_short}
\end{subfigure}
\caption{\footnotesize (a) Conditional entropy for communication using 1 and 5-line sketches with two target contexts (same or different between the 2 agents). (b) Communication accuracy on two altered numerosity datasets with an unequal number of training samples: increasing from class 1, having the fewest samples, to class 5, having the most, and vice versa. (c) Test sketches from training on the 3 different datasets: equal, increasing and decreasing number of class samples.}
\label{fig:sketching_freq_accuracies}
\end{figure}

\paragraph{Communication success and code precision}
First, we observe that agents can reliably communicate trained numerosities, even in the \emph{Diff} condition, where the sender and receiver see different instances of the same numerosity (Fig.~\ref{subfig:pointplot_accuracy_same_v_diff} and \ref{subfig:pointplot_sketch_acc_same_vs_diff}). The system becomes more precise in the different setups indicated by the lower entropy (Fig.~\ref{subfig:pointplot_conditional_e_same_v_diff} and \ref{subfig:point_plot_conditional_entropy_same_v_diff_1v5_strokes}). Under length regularisation (Fig.~\ref{fig:discrete-regularization}), we observe that messages become shorter with minimal accuracy loss and lower entropy, indicating the pressure toward efficient bijective encodings. 
This shows that even without language-internal conventions, precision emerges from the joint pressure of coordination and information bottleneck.

\vspace{-1em}

\vspace{-2.3em}

\vspace{-2em}

\paragraph{Arbitrariness outweighs frequency}
We next vary the training context to assess whether more frequently occurring numerosities are encoded more economically compared to the least frequent ones. We find that manipulating class frequency (Uniform/Increase/Decrease setups detailed in supplementary) barely changes accuracy (Fig.~\ref{subfig:point_plot_accucary_increase_v_decrease}) and does not induce shorter codes for frequent numerosities. The sketch channel tends to scale with the spatial span of the input image elements, rather than with frequency (Fig.~\ref{subfig:sketch_examples_inc_dec_short}). Human numerals often show mixed arbitrariness with some high-frequency lexicalisation. In this set of experiments, the absence of transmission costs beyond accuracy likely weakens Zipf-like simplification, and the arbitrariness survives because it is \textit{good enough} for coordination.

\vspace{-1em}

\begin{figure}[htb]
\centering
\begin{subfigure}{0.49\linewidth}
    \includegraphics[width=\linewidth]{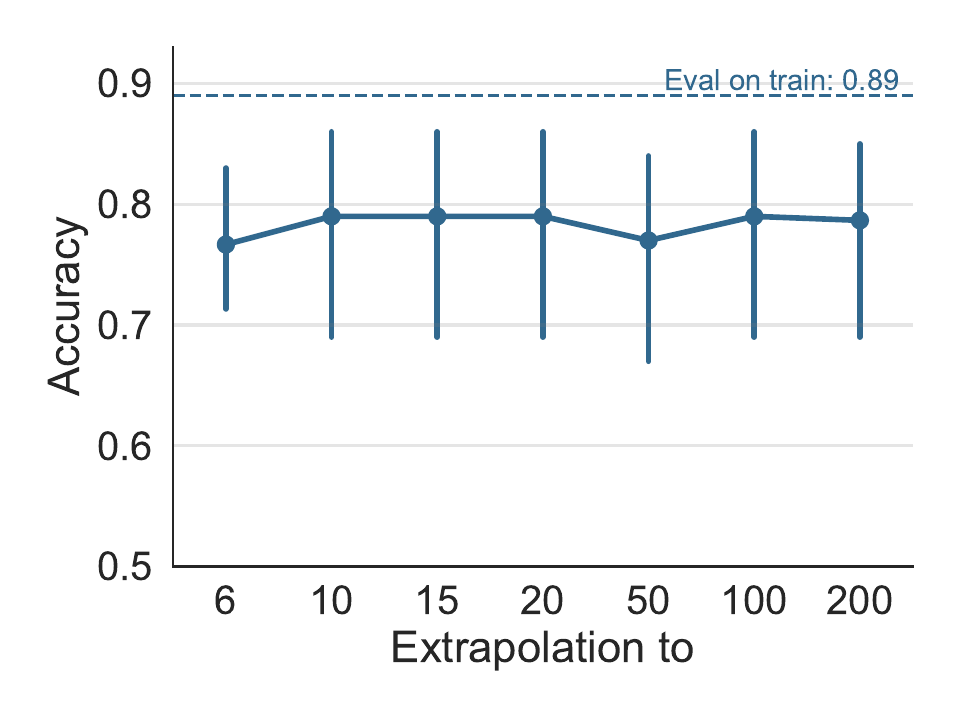}
    \phantomsubcaption
    \label{subfig:pointplot_accuracy_extrapolations}
\end{subfigure}
\begin{subfigure}{0.49\linewidth}
    \includegraphics[width=\linewidth]{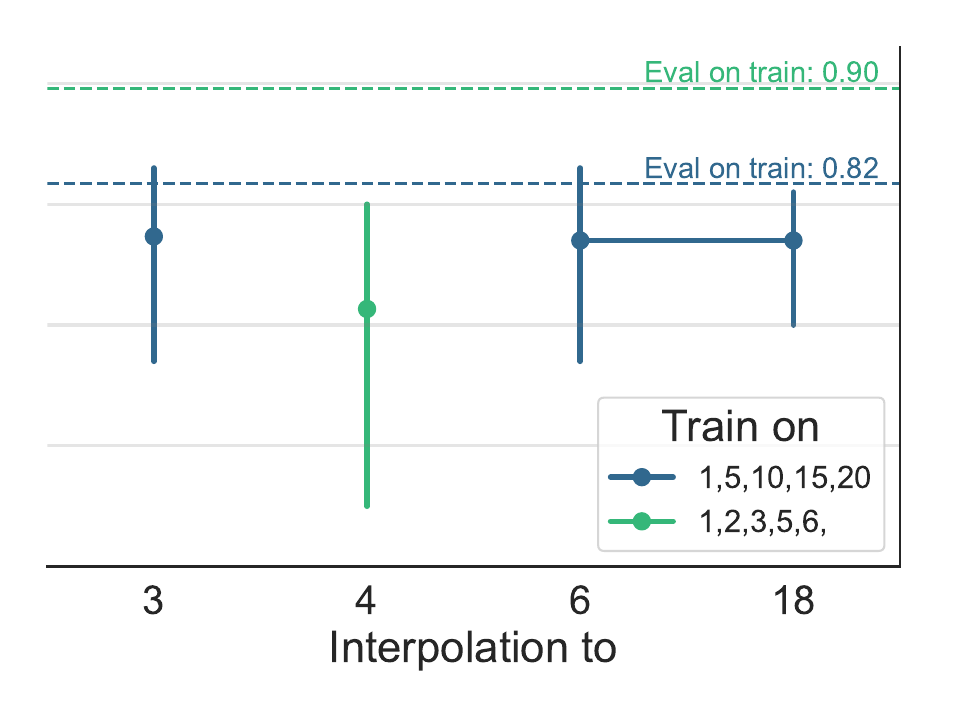}
    \phantomsubcaption
    \label{subfig:pointplot_accuracy_interpolations}
\end{subfigure}\\
\vspace{-1.2em}
\begin{subfigure}{1\linewidth}
    \includegraphics[width=\linewidth]{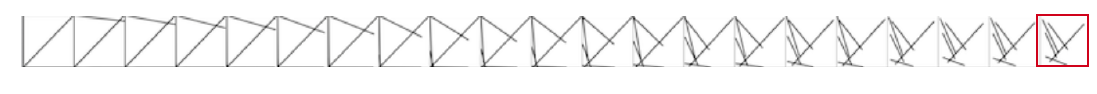}
    \phantomsubcaption
    \label{subfig:sketches_extrapol}
\end{subfigure}
\vspace{-3em}
\caption{\footnotesize (Left) \textbf{Discrete extrapolation}: model trained on $\{1,2,3,4,5\}$. Test communication accuracy is reported for each test set (train $+$ one out-of-distribution numerosity). The model uses the same message that was previously chosen for the highest pretrained numerosity (\eg 5) for every new higher extrapolation. (Right) \textbf{Interpolation} results from two different training sets. (Bottom) Example of sketch extrapolations from training within $\{1-20\}$ to \emph{25} highlighted in red. }
\label{fig:discrete-extrapolations}
\end{figure}

\vspace{-2.2em}
\paragraph{Generalisation and compositionality} We further investigate whether the code developed by the agents can generalise to unseen numerosities (Discrete setup shown in Fig.~\ref{fig:discrete-extrapolations} Top). We find that under out-of-distribution testing, agents typically reuse the message or symbol for the largest trained numerosity. 
The interpolation analysis, testing within-the-range numerosities, also indicates a lack of systematic structure. Sketch-based communication behaves similarly, although on average, extrapolations with a continuous channel are better than chance. 
However, agents do not develop a compositional sketch but rather map everything larger than in training onto a single sketch, which usually resembles the highest seen numerosity (\eg  Fig.~\ref{fig:discrete-extrapolations}-Bottom, and supplementary).
The emerging signals stabilise as holistic labels, with a separate message/sketch for each trained numerosity and no subparts are reused to compose new messages. 


\section{Implications for Language Evolution}

\paragraph{Coordination yields precision and numerical representations.} The agents are able to quickly learn low-entropy codes, precisely communicating about trained numerosities. Numerosities are communicated independently of visual features, indicating the acquisition of abstract numerical representations.

\vspace{-0.7em}

\paragraph{Iconicity helps, but arbitrariness wins.} The sketch communication channel shows some size and span cues, with sketch size capturing the spread of the dots, rather than the quantity; but both discrete and continuous communication channels settle into mostly arbitrary, class-specific signals. 

\vspace{-0.7em}

\paragraph{Frequency alone doesn't yield efficient codes.} Human languages often make frequent words shorter, but here, without explicit costs for longer messages, skewed input distributions do not result in more efficient codes. Adding production and perception costs could aid the brevity of the emergent codes.

\vspace{-0.7em}

\paragraph{Lack of compositionality, lack of generalisation}. 
When faced with higher, untrained numerosities, agents reuse the message associated with the largest trained numerosity. This limitation may relate to the lack of compositionality in the emergent communication system: agents use holistic messages, and neither reuse interpretable message parts nor develop rule-based compositions. While holistic messages encoding non-decomposable bundles of features commonly emerge in referential games~\cite{lacroix2025information,freeborn2025compositional}, compositionality plays a crucial role in bootstrapping numerical representations, as evidenced by human learning. Inducing such compositionality in this context remains a key challenge. Unlike referents defined by multiple features (\eg, colour and shape), numbers vary along a single ordered dimension, offering no inherently decomposable features to encode.  Successful generalisation, therefore, requires the emergence of an arbitrary but systematic partitioning of this dimension, rather than the transparent mapping of perceptual features. How such a structure can be induced remains an open question. Alternative learning setups may help, such as population training \shortcite<e.g.>{mahaut2025referential} or other mechanisms that promote compositionality in both human communication and neural networks, such as iterated learning, vocabulary constraints, and biases toward symbol reuse~\cite{lake2023human,kirby2008cumulative}. 
Exploring such mechanisms may therefore be key to enabling systematic numerical generalisation. 

\section*{Acknowledgements}
FF received funding from the European Research Council (ERC) under the European Union’s Horizon 2020 research and innovation program (grant agreement No.~101019291) and from the Catalan government (AGAUR grant SGR 2021 00470).

\bibliographystyle{apacite}
\bibliography{evolang}

@article{barner2017language,
  title={Language, procedures, and the non-perceptual origin of number word meanings},
  author={Barner, David},
  journal={Journal of child language},
  volume={44},
  number={3},
  pages={553--590},
  year={2017},
  publisher={Cambridge University Press}
}

@book{overmann2025cultural,
  title={Cultural number systems: A sourcebook},
  author={Overmann, Karenleigh A},
  year={2025},
  publisher={Springer Nature}
}

@article{lacroix2025information,
  title={Information and Meaning in the Evolution of Compositional Signals: T. LaCroix},
  author={LaCroix, Travis},
  journal={Journal of Logic, Language and Information},
  pages={1--21},
  year={2025},
  publisher={Springer}
}

@article{freeborn2025compositional,
  title={Compositional understanding in signaling games},
  author={Freeborn, David Peter Wallis},
  journal={Synthese},
  volume={206},
  number={3},
  pages={116},
  year={2025},
  publisher={Springer}
}

@book{menninger2013number,
  title={Number words and number symbols: A cultural history of numbers},
  author={Menninger, Karl},
  year={2013},
  publisher={Courier Corporation}
}

@article{mahaut2025referential,
title={Referential communication in heterogeneous communities of pre-trained visual deep networks},
author={Mat{'e}o Mahaut and Roberto Dessi and Francesca Franzon and Marco Baroni},
journal={Transactions on Machine Learning Research},
issn={2835-8856},
year={2025},
url={https://openreview.net/forum?id=8L3khbpUJL},
note={}
}

@incollection{wals-131,
  author    = {Bernard Comrie},
  booktitle = {The World Atlas of Language Structures Online},
  editor    = {Matthew S. Dryer and Martin Haspelmath},
  publisher = {Zenodo},
  title     = {Numeral Bases (v2020.3)},
  type      = {Data set},
  url       = {https://doi.org/10.5281/zenodo.7385533},
  year      = {2013},
  doi       = {10.5281/zenodo.7385533}
}

@article{coolidge2012numerosity,
  title={Numerosity, abstraction, and the emergence of symbolic thinking},
  author={Coolidge, Frederick L and Overmann, Karenleigh A},
  journal={Current Anthropology},
  volume={53},
  number={2},
  pages={204--225},
  year={2012},
  publisher={University of Chicago Press Chicago, IL}
}

@article{denic2024recursive,
  title={Recursive Numeral Systems Optimize the Trade-off Between Lexicon Size and Average Morphosyntactic Complexity},
  author={Deni{\'c}, Milica and Szymanik, Jakub},
  journal={Cognitive Science},
  volume={48},
  number={3},
  pages={e13424},
  year={2024},
  publisher={Wiley Online Library}
}

@article{pajot2025compositional,
  title={The compositional nature of number concepts: Insights from number frequencies},
  author={Pajot, Maxence and Sabl{\'e}-Meyer, Mathias and Dehaene, Stanislas},
  journal={Cognition},
  volume={263},
  pages={106213},
  year={2025},
  publisher={Elsevier}
}

@article{xu2020numeral,
  title={Numeral systems across languages support efficient communication: From approximate numerosity to recursion},
  author={Xu, Yang and Liu, Emmy and Regier, Terry},
  journal={Open Mind},
  volume={4},
  pages={57--70},
  year={2020},
  publisher={MIT Press One Rogers Street, Cambridge, MA 02142-1209, USA journals-info~…}
}

@inproceedings{mihai2021learningtodraw,
 author = {Mihai, Daniela and Hare, Jonathon},
 booktitle = {Advances in Neural Information Processing Systems},
 editor = {M. Ranzato and A. Beygelzimer and Y. Dauphin and P.S. Liang and J. Wortman Vaughan},
 pages = {7153--7166},
 publisher = {Curran Associates, Inc.},
 title = {Learning to Draw: Emergent Communication through Sketching},
 url = {https://proceedings.neurips.cc/paper_files/paper/2021/file/39d0a8908fbe6c18039ea8227f827023-Paper.pdf},
 volume = {34},
 year = {2021}
}

@inproceedings{radford2021learning,
  title = 	 {Learning Transferable Visual Models From Natural Language Supervision},
  author =       {Radford, Alec and Kim, Jong Wook and Hallacy, Chris and Ramesh, Aditya and Goh, Gabriel and Agarwal, Sandhini and Sastry, Girish and Askell, Amanda and Mishkin, Pamela and Clark, Jack and Krueger, Gretchen and Sutskever, Ilya},
  booktitle = 	 {Proceedings of the 38th International Conference on Machine Learning},
  pages = 	 {8748--8763},
  year = 	 {2021},
  editor = 	 {Meila, Marina and Zhang, Tong},
  volume = 	 {139},
  series = 	 {Proceedings of Machine Learning Research},
  month = 	 {18--24 Jul},
  publisher =    {PMLR},
  url = 	 {https://proceedings.mlr.press/v139/radford21a.html},
}

@article{ReviewEmeComm,
title={A Review of the Applications of Deep Learning-Based Emergent Communication},
author={Brendon Boldt and David R Mortensen},
journal={Transactions on Machine Learning Research},
issn={2835-8856},
year={2024},
url={https://openreview.net/forum?id=jesKcQxQ7j},
note={}
}

@article{hochreiter1997long,
  title={Long short-term memory},
  author={Hochreiter, Sepp and Schmidhuber, J{\"u}rgen},
  journal={Neural computation},
  volume={9},
  number={8},
  pages={1735--1780},
  year={1997},
  publisher={MIT press}
}

@article{brandizzi2023toward,
  title={Toward more human-like ai communication: A review of emergent communication research},
  author={Brandizzi, Nicolo’},
  journal={IEEE Access},
  volume={11},
  pages={142317--142340},
  year={2023},
  publisher={IEEE}
}

@article{kirby2008cumulative,
  title={Cumulative cultural evolution in the laboratory: An experimental approach to the origins of structure in human language},
  author={Kirby, Simon and Cornish, Hannah and Smith, Kenny},
  journal={Proceedings of the National Academy of Sciences},
  volume={105},
  number={31},
  pages={10681--10686},
  year={2008},
  publisher={National Academy of Sciences}
}

@article{lake2023human,
  title={Human-like systematic generalization through a meta-learning neural network},
  author={Lake, Brenden M and Baroni, Marco},
  journal={Nature},
  volume={623},
  number={7985},
  pages={115--121},
  year={2023},
  publisher={Nature Publishing Group UK London}
}

@article{lewis1969convention,
  title={Convention: A Philosophical Study},
  author={Lewis, David Kellogg},
  year={1969}
}
\clearpage
\appendix

\begin{center}
    {\bfseries Supplementary Material}
\end{center}


\section{Manipulating Class Frequency - Experimental Setup}
\label{supp-sec:class-frequency}

We vary the training context to assess whether more frequent, and thus varied, numerosities trigger a more efficient code, and the least varied correspond to a more complex code. 
While we use a \textit{Uniform} distribution in most of our experiments and with an equal number of training samples for each numerosity (700), we also define \textit{Increase} and \textit{Decrease} setups. Therefore, we define \textit{Increase} setup to have the following distribution of samples per class: $100,200,300,400,700$ increasing from numerosity 1 to numerosity 5; and the reverse for the \textit{Decrease} setup. 

\begin{figure}[ht]
\centering
    \includegraphics[width=0.44\linewidth]{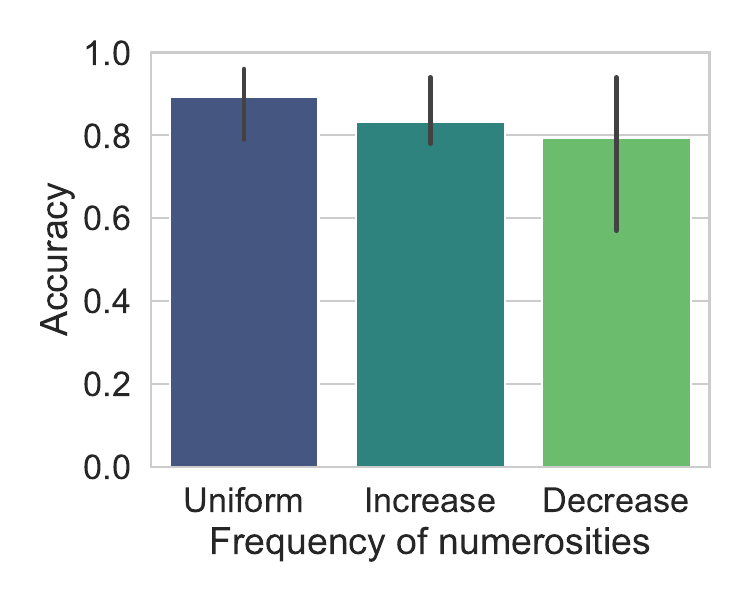}
    \label{subfig:freq-discrete-accuracy}
\caption{\footnotesize Accuracy and standard deviation across 3 seeds for communication task performed with different frequencies of training numerosity samples: \textit{Uniform}, \textit{Increase} and \textit{Decrease} frequency setups.}
\label{fig:frequency-discrete-accuracy}
\end{figure}

\begin{figure*}[htb]
\centering
\begin{subfigure}{0.32\linewidth}
    \includegraphics[width=\linewidth]{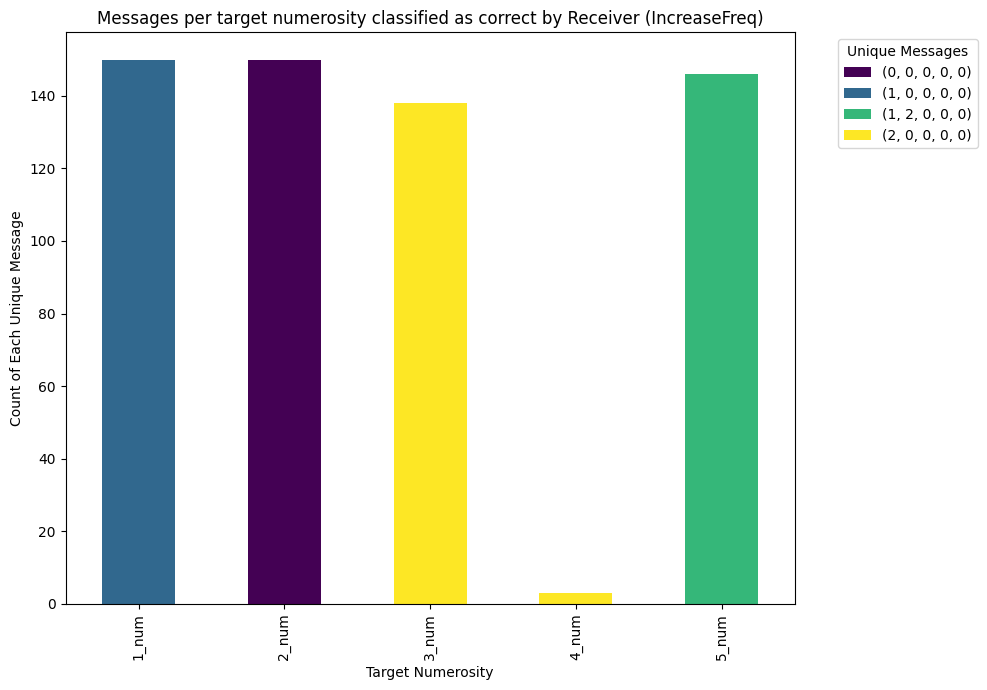}
    \caption{seed 0}
    \label{subfig:freq-discrete-increase-seed0}
\end{subfigure}
\hfill
\begin{subfigure}{0.32\linewidth}
    \includegraphics[width=\linewidth]{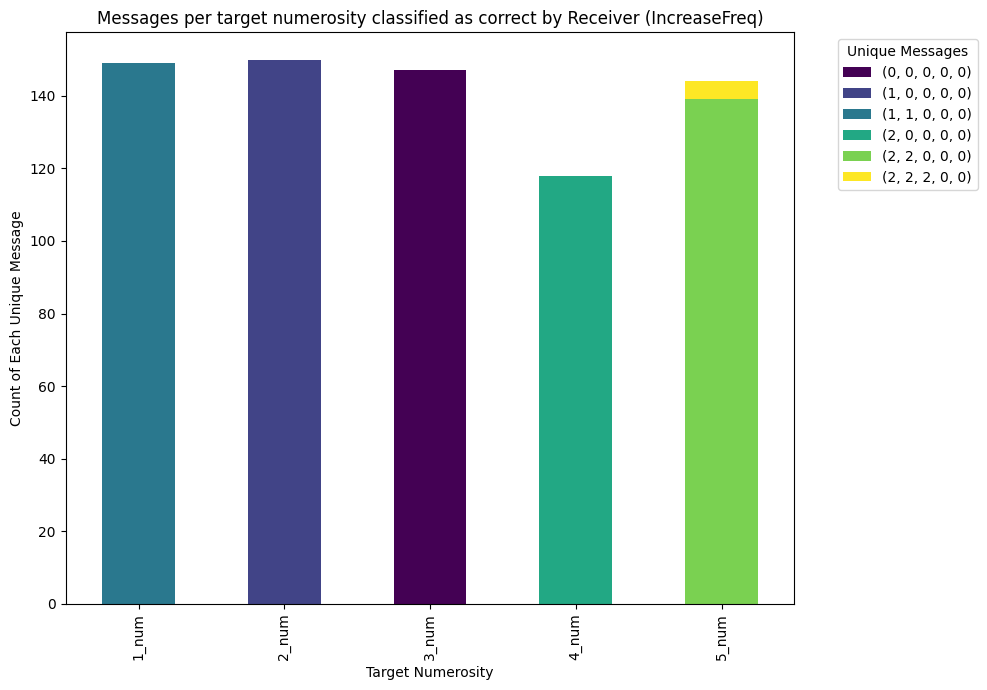}
    \caption{seed 1}
    \label{subfig:freq-discrete-increase-seed1}
\end{subfigure}
\hfill
\begin{subfigure}{0.32\linewidth}
    \includegraphics[width=\linewidth]{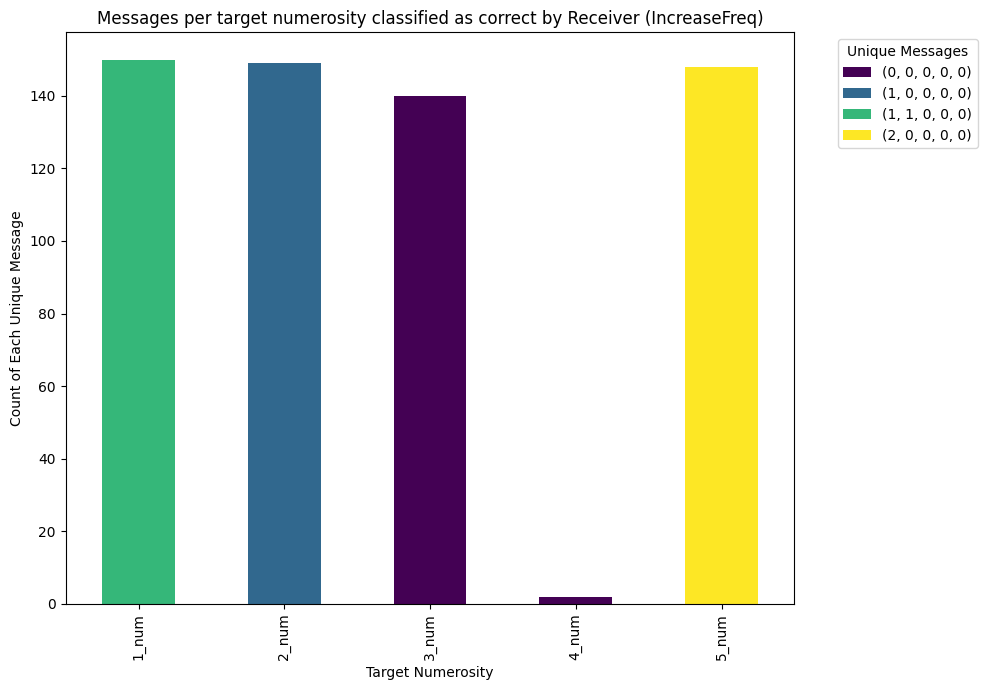}
    \caption{seed 2}
    \label{subfig:freq-discrete-increase-seed2}
\end{subfigure}
\hfill
\begin{subfigure}{0.32\linewidth}
    \includegraphics[width=\linewidth]{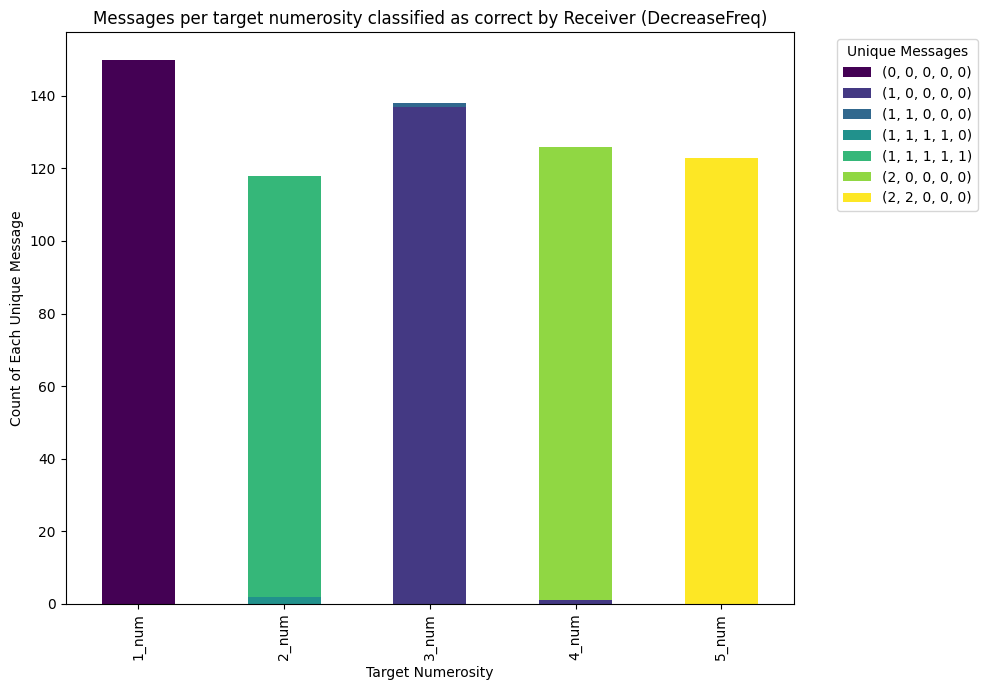}
    \caption{seed 0}
    \label{subfig:freq-discrete-decrease-seed0}
\end{subfigure}
\hfill
\begin{subfigure}{0.32\linewidth}
    \includegraphics[width=\linewidth]{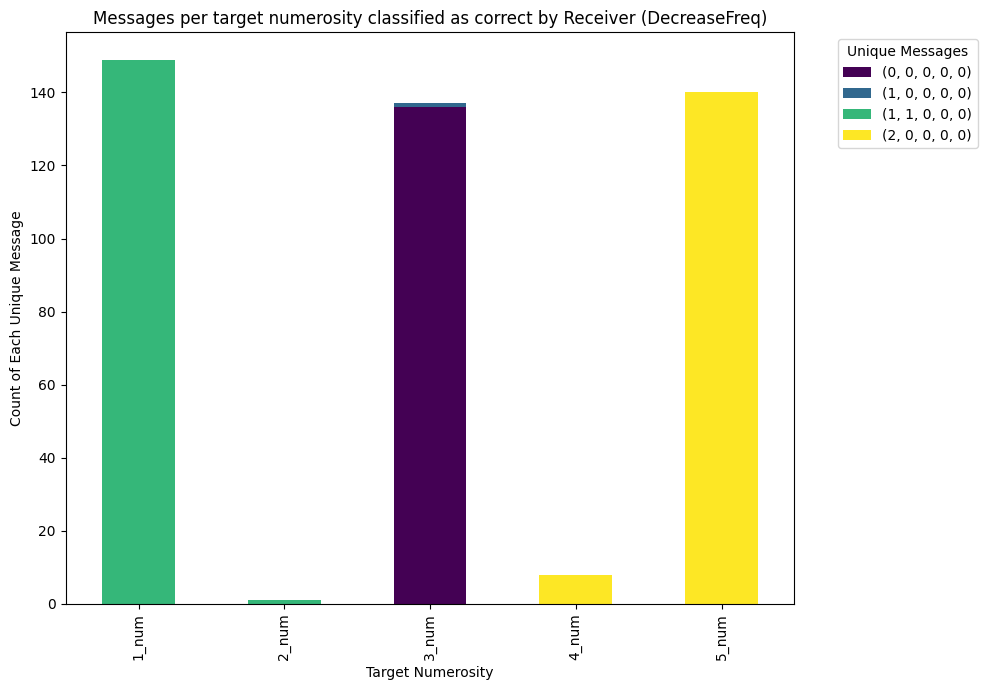}
    \caption{seed 1}
    \label{subfig:freq-discrete-decrease-seed1}
\end{subfigure}
\hfill
\begin{subfigure}{0.32\linewidth}
    \includegraphics[width=\linewidth]{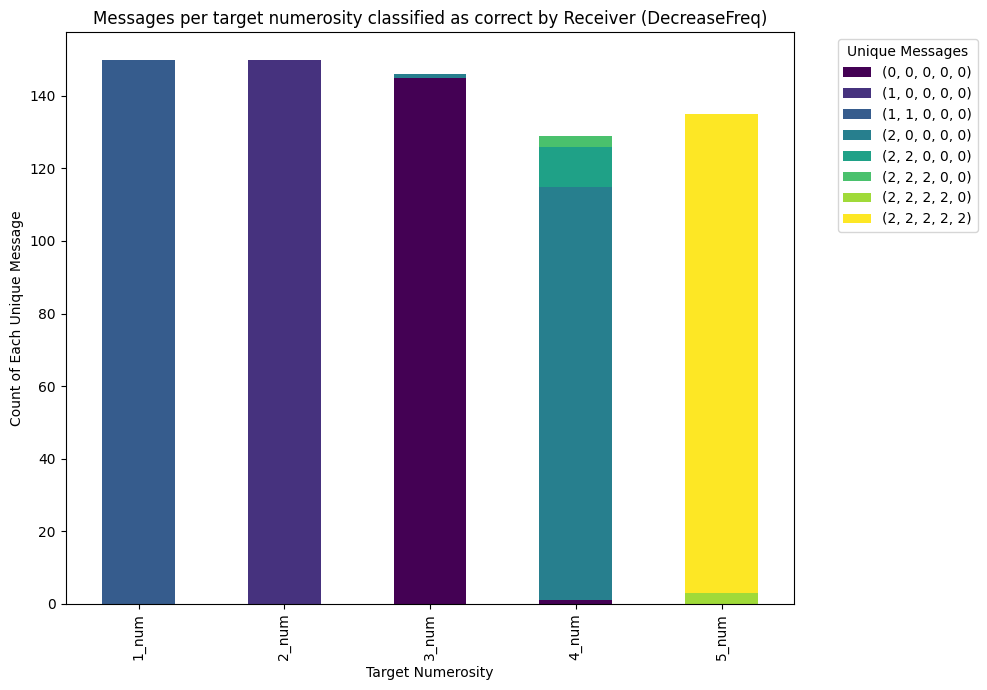}
    \caption{seed 2}
    \label{subfig:freq-discrete-decrease-seed2}
\end{subfigure}
\caption{\footnotesize Exact message to numerosity mappings for correctly classified numerosities, across 3 different seeds, in the \textit{Increase} (a, b and c) and \textit{Decrease} frequency setups (d, e, f).}
\label{fig:frequency-exact-mappings-discrete}
\end{figure*}

The model used for this experiment has variable message length, up to 5 tokens, a vocabulary of size 3 and a message regularisation loss with a coefficient of 0.005. \fref{fig:frequency-discrete-accuracy} shows the discrete channel test accuracies for different training data distributions, and \fref{fig:frequency-exact-mappings-discrete} illustrates the exact mappings for the \textit{Increase} and \textit{Decrease} setups across 3 seeds.

\begin{figure*}[hb]
\begin{subfigure}{0.32\linewidth}
    \includegraphics[width=\linewidth]{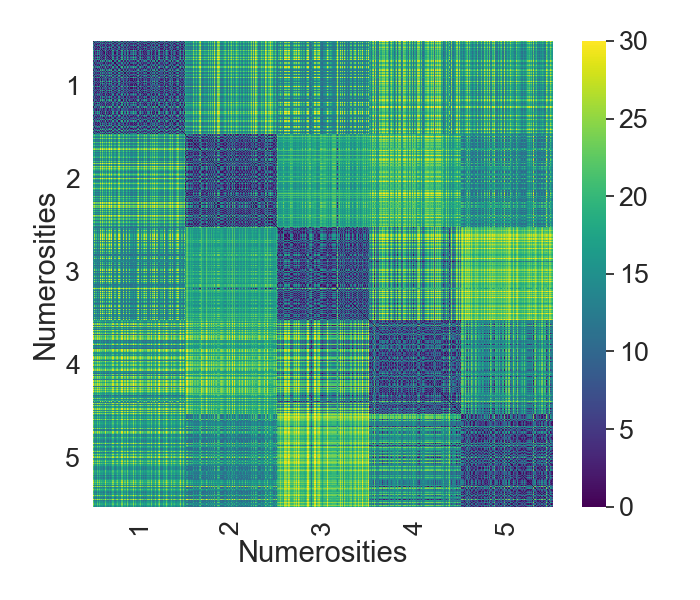}
    \vspace{-1.5\baselineskip}
    \caption{Uniform}
    \label{subfig:sketch_disimilarity_equal}
\end{subfigure}
\hfill
\begin{subfigure}{0.32\linewidth} 
    \includegraphics[width=\linewidth]{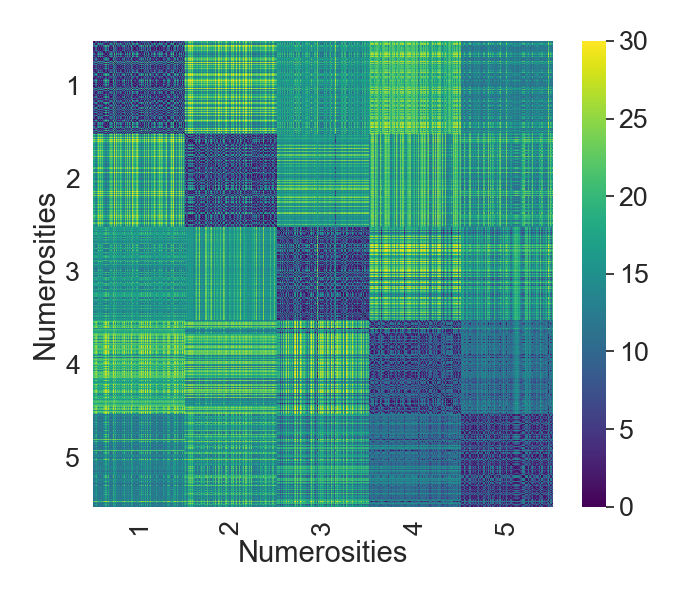}
    \vspace{-1.5\baselineskip}
    \caption{Increase}
    \label{subfig:sketch_disimilarity_increase}
\end{subfigure}
\hfill
\begin{subfigure}{0.32\linewidth} 
    \includegraphics[width=\linewidth]{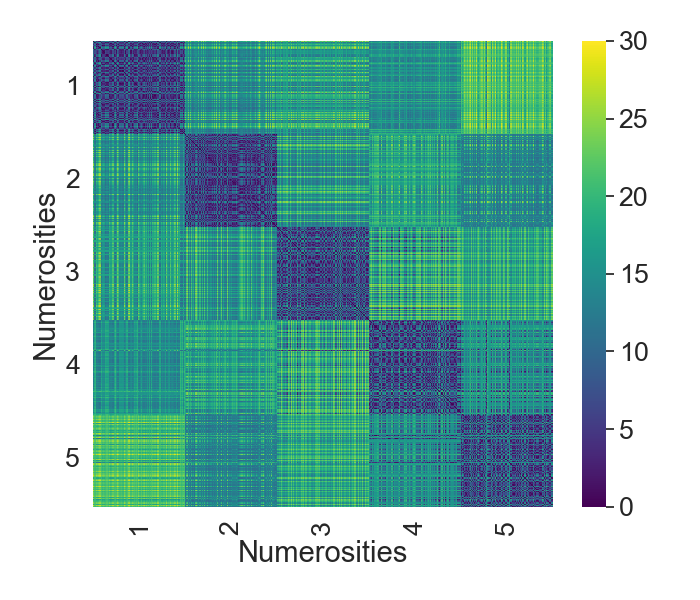}
    \vspace{-1.5\baselineskip}
    \caption{Decrease}
    \label{subfig:sketch_disimilarity_decrease}
\end{subfigure}
\hfill
\caption{\footnotesize Pair-wise dissimilarity for sketches in the tSNE-reduced pixel-space for various frequencies of numerosities: (a) equal, (b) increasing from 1 to 5, (c) decreasing from 1 to 5.}
\label{fig:heatmaps_frequency_sketches}
\end{figure*}

The continuous channel accuracies across seeds are reported in Figure 4b of the main paper and are similar in trend to those of the discrete channel (\fref{fig:frequency-discrete-accuracy}). Further examples of sketches produced at test time for each of the numerosity classes present in the training set are shown in \fref{fig:sketch_num_inc_decrease_full}. The different rows indicate results from agents being trained with more or less diverse training samples, some classes having more or fewer instances. The pairwise dissimilarity of sketches in the three varying frequency setups is shown in \fref{fig:heatmaps_frequency_sketches}.

The results indicate the messages are arbitrary and there is no concrete correlation between message complexity and the number of training samples. The only consistent trend is observed in the accuracy across the different training setups, \textit{Decreasing} being the most difficult to solve with both a discrete and continuous channel. 

An observation from this experiment is that while the symbols do not vary in complexity based on the frequency, nor are they compositional (in the sense of combining more frequent numerosities' codes), they all seem to correlate with the area filled by elements. For example, numerosity 1 is usually represented through a symbol whose lines are concentrated at the centre of the sketch and do not cover much of the available space. As the numerosity increases and the elements start spreading across the target image space, the lines start increasing in length and are positioned in such a way as to cover as much of the available sketching space.

\clearpage
\section{Extended Results}\label{app:Results}

\begin{table}[hb]
    \centering
    \caption{\footnotesize Model with fixed message length trained on numerosities $\{1,2,3,4,5\}$ with different $vocabulary$ sizes: $\{3,5,10,100\}$. Test communication accuracy and entropy are across 3 seeds.}
    \begin{tabular}{lcccc}
        \toprule
        {\textbf{Setup}}& 
        {\textbf{Max Msg Length}} & {\textbf{Vocabulary Size}}& {\textbf{Accuracy}}& 
        \textbf{Entropy}  \\
        \midrule
        \multirow{4}{*}{Diff target} & 
        \normalsize{5} &
        \normalsize{3} &
        \normalsize{.95}\scriptsize{$\pm$.2}& \normalsize{1.14}\scriptsize{$\pm$.34}\\
        & \normalsize{5} &  
        \normalsize{5} & \normalsize{.96}\scriptsize{$\pm$0}& \normalsize{1.15}\scriptsize{$\pm$.31}\\
        & \normalsize{5} &
        \normalsize{10} & \normalsize{.94}\scriptsize{$\pm$.02}& \normalsize{1.13}\scriptsize{$\pm$.42}\\
        & \normalsize{5} &
        \normalsize{100} & \normalsize{.96}\scriptsize{$\pm$.008}& \normalsize{0.76}\scriptsize{$\pm$.15}\\
        \bottomrule
       \end{tabular}
    \label{tab:vocabsize-discrete}
\end{table} 

\begin{figure}[hb]
\centering
\begin{subfigure}{0.32\linewidth}
    \includegraphics[width=\linewidth]{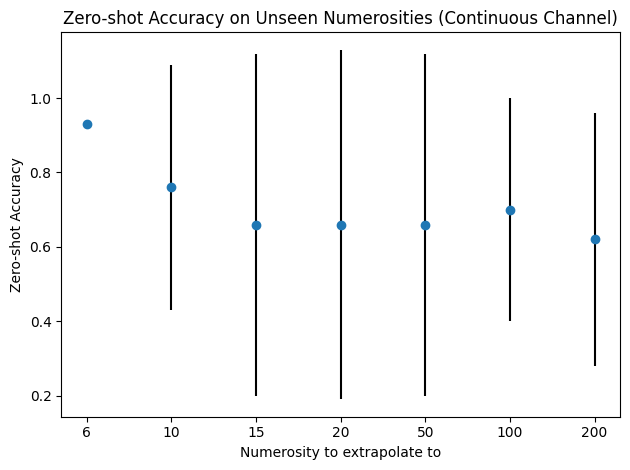}
\caption{Extrapolation}
\label{subfig:zeroshot-extrapol-continuous}
\end{subfigure}
\hfill
\begin{subfigure}{0.32\linewidth}
    \includegraphics[width=\linewidth]{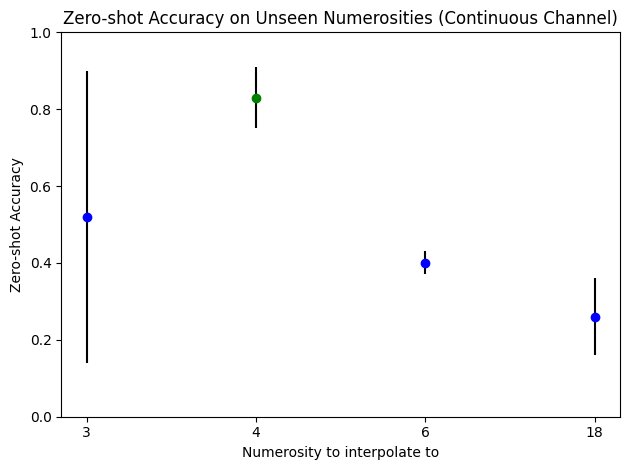}
\caption{Interpolation}
\label{subfig:zeroshot-interpol-continuous}
\end{subfigure}
\hfill
\begin{subfigure}{0.32\linewidth}
    \includegraphics[width=\linewidth]{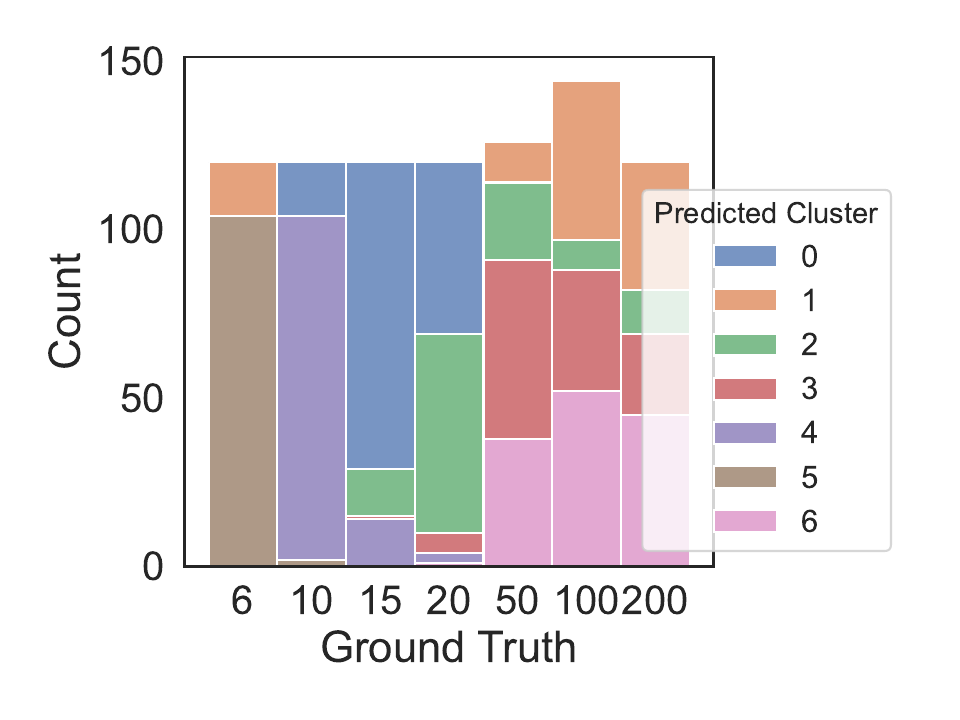}
\caption{Mapping of sketches}
\label{subfig:sketching_matching_numerosity_to_message}
\end{subfigure}
\caption{\footnotesize (a, b) Zero-shot accuracies for out-of-distribution testing data for the continuous (\emph{i.e.} sketching) channel. (a) Extrapolation: model trained on \emph{train=$\{1, 2, 3, 4, 5\}$} and tested on \emph{train} $+$ one OOD. (b) Interpolation: the different colours in interpolation (green vs blue) correspond to different training data - please refer to Fig. 5 in the main paper. c) Mapping of sketches after clustering based on similarity to extrapolated numerosities (cond. entropy: 1.348).}
\label{fig:sketch_zeroshot-accuracies}
\end{figure}

\begin{figure*}[htb]
\centering
\begin{subfigure}{0.7\linewidth}
    \includegraphics[width=\linewidth]{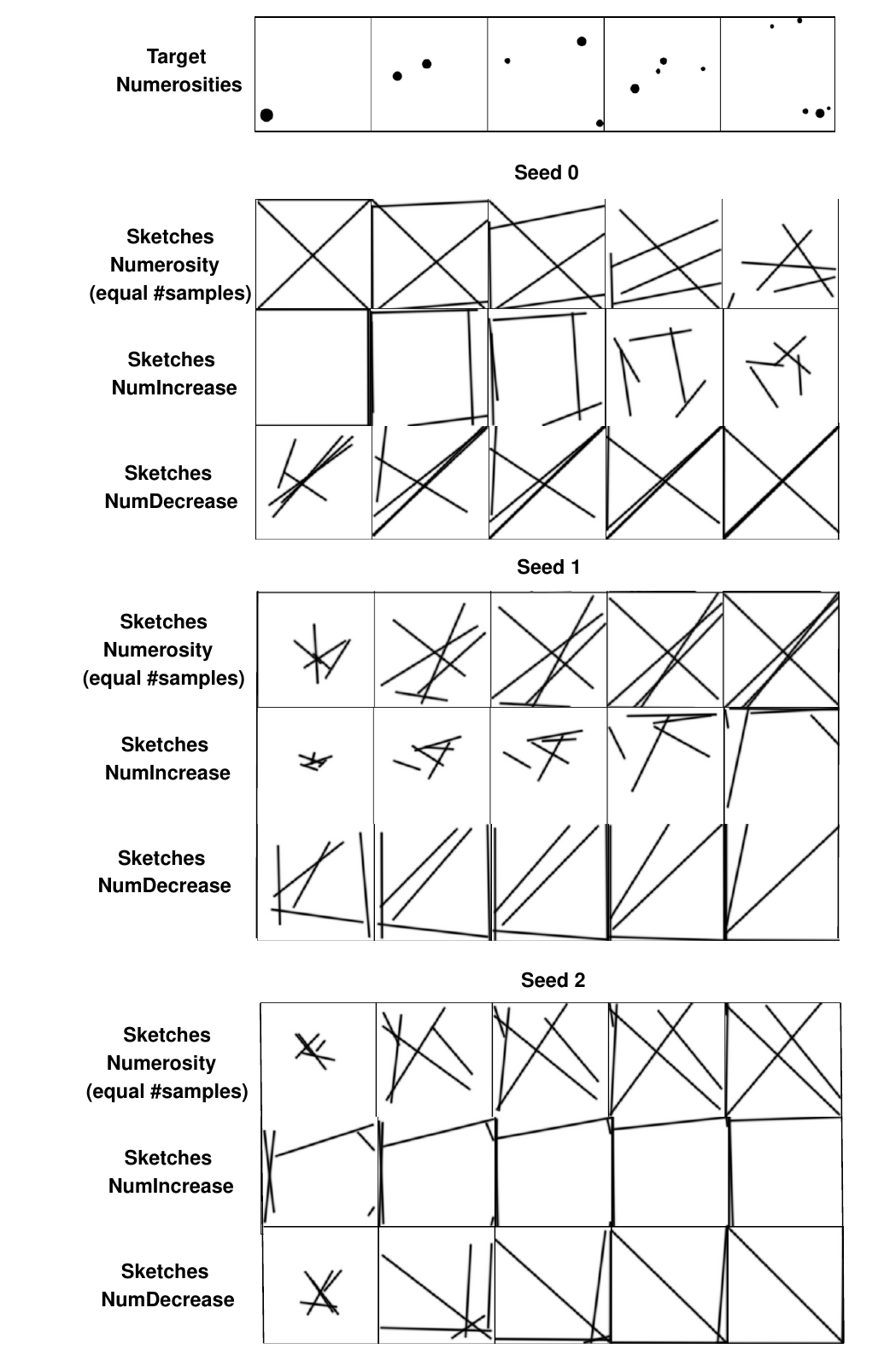}
\end{subfigure}
\caption{\footnotesize Test target images (each entry in the first row corresponds to a distinct numerosity between 1 and 5) and sketches produced by agents trained on different numerosity datasets, with varying numbers of class samples. Results are reported across 3 training seeds.}
\label{fig:sketch_num_inc_decrease_full}
\end{figure*}

\end{document}